\documentclass[sensors,article,accept,pdftex,moreauthors]{Definitions/mdpi}

\usepackage{environ}
\NewEnviron{myequation}{%
\begin{equation}
\scalebox{0.85}{$\BODY$}
\end{equation}
}

\firstpage{1} 
\pubvolume{1}
\issuenum{1}
\articlenumber{0}
\pubyear{2022}
\copyrightyear{2022}
\externaleditor{Academic Editor: Changchuan Yin,  Mingzhe Chen and Zhaohui Yang}  

\datereceived{7 March 2022} 
\dateaccepted{25 March 2022} 
\datepublished{} 
\hreflink{https://doi.org/} 

\usepackage{algorithm}
\usepackage{algorithmic}
\usepackage{bm}

\Title{Precoder and Decoder Co-Designs for Radar and Communication Spectrum Sharing $^\dagger$}

\TitleCitation{Precoder and Decoder Co-Designs for Radar and Communication Spectrum Sharing}

\Author{{Yuanhao Cui} $^{1}$, Xiaojun Jing $^{1,}$* and Visa Koivunen $^{2}$}

\AuthorNames{Yuanhao Cui, Xiaojun Jing and Visa Koivunen}

\AuthorCitation{Cui, Y.; Jing, X.; Koivunen, V.}

\address{%
$^{1}$ \quad {School of Information and Communication Engineering}, Beijing University of Posts and Telecommunications, Beijing {100876}, China; \{cuiyuanhao, jxiaojun\}@bupt.edu.cn\\
$^{2}$ \quad Department of Signal Processing and Acoustics, Aalto University, {00076 Aalto}, Finland; visa.koivunen@aalto.fi 
}

\corres{Correspondence: jxiaojun@bupt.edu.cn}

\firstnote{This paper is an extended version of our paper published in Cui, Y.; {Koivunen, V.}; Jing, X. Interference alignment based precoder--decoder design for radar-communication co-existence. In the Proceedings of the 51st Asilomar Conference on Signals, Systems, and Computers, Pacific Grove, CA, USA, 29 October--1 November 2017; pp.~1290--1295.} 

\abstract{{Radar and modern communication systems are both evaluating towards higher frequency bands and massive antenna arrays, thus increasing their similarities in terms of hardware structure, channel characteristics, and signal processing pipelines. To suppress the cross-system interference caused by communications and radar systems with shared spectral and hardware resources, the co-design philosophy, wherein the communications and radar/sensing systems can operate in parallel with jointly optimized performance, has drawn substantial attention from both academia and industry.} 
In this paper, we propose a nullspace-based joint precoder-decoder design for spectrum sharing between multicarrier radar and multiuser multicarrier communication systems, by employing the maximizing signal interference noise ratio (max-SINR) criterion and interference alignment (IA) constraints. By projecting the cross-system interference to the designed null spaces, a maximum degree of freedom upper bound for the $K+1$-radar-communication-user interference channel can be achieved.
Our simulation studies demonstrate that interference can be practically fully canceled in both communication and radar systems. This leads to improved detection performance in radar and a higher rate in communication subsystems. A significant performance gain over a nullspace-based precoder-only design \cite{mahal2017spectral} is also obtained.}


\begin{document}

\section{Introduction}

\subsection{Background and Motivation}

The tremendous growth in new wireless systems and services has caused a shortage in the useful radio spectrum~\cite{2012president}.
For instance, the Federal Communications Commission (FCC)
and the National Telecommunications and Information Administration (NTIA) proposed
to adjust the allocation of 150 megahertz bandwidth from surveillance and air defense to
 radar and communication systems in the 3.5 GHz band~\cite{federal2012enabling,locke2010assessment,liu2021integrated,cui2021integrating}. Agile spectrum use and spectrum sharing among different radio systems provide promising technologies for alleviating the lack of useful spectrum. 
 
 By exploiting radar spatial degrees of freedom (DoFs), the joint transmit design of a multiple-input multiple-output matrix completion (MIMO-MC) radar and a point-to-point multiple-input multiple-output (MIMO) communication system~\cite{li2016joint,li2017joint} has received considerable  {attention}. {However, the influence of cross-system interference in} {joint radar-communication scenarios} {is still not thoroughly studied.}
 
 In terms of spectrum sharing in the 3.5 GHz spectrum and hardware similarity, a multicarrier radar waveform is considered to be one of the best choices for spectrum sharing between radar and the communication system. Indeed, a multicarrier {waveform} has already been widely accepted as a physical layer modulation solution both in communication systems~\cite{3gpp.36.331, IEEE2001,IEEE201} and in the radar field~\cite{levanon2000multifrequency,levanon2002multicarrier,sen2011multiobjective,sen2011adaptive}. {Furthermore, a communication signal that is scattered or reflected by the target may also be {utilized} for radar sensing purposes \cite{bica2016mutual,shi2017power}.
}
Currently, the interference signal can be projected into the nullspace of the channel matrix to avoid interference {at the receiver}. The majority of recent projection-based radar-communication coexistence techniques, e.g.,~\cite{sodagari2012projection,mahal2017spectral,khawar2014beampattern}, project either radar or communication signals into the other system's nullspace. {Among them\cite{sodagari2012projection,mahal2017spectral,khawar2014beampattern}, a multiuser communication system is simplified to a communication subsystem with combined signal spaces of all communication users, ignoring their mutual interference. Then, the radar subsystem searches for a nullspace or alternative nullspace of the communication subsystem 
into which the cross-system interference is projected.}

However, the feasibility of the {precoder-only design} depends on the {availability of} channel state matrix, {and on the high-quality feedback from the receiver to the transmitter.} {Furthermore, comparing with precoder-only projection-based design~\cite{sodagari2012projection,mahal2017spectral,khawar2014beampattern}, this type of
precoder design only allows for avoiding interference on each subsystem but not for both.}

In the case that radar and communication systems are jointly designed and co-located, {the sharing of channel information and interference awareness can be easily arranged. Consequently, the transmitted waveforms and receiver processing for radar and communication systems can be jointly optimized.} For example, precoders and decoders in radar and communication systems may be jointly designed to construct signal spaces and orthogonal interference spaces and to obtain more effective and flexible interference management. 

\subsection{Contributions of This Work}

In this paper, we propose  {interference alignment (IA) ~\cite{cadambe2008interference} based precoder and decoder co-design to manage interference for spectrum sharing between multicarrier radar~\cite{bica2016generalized} communication systems}. DoF exploitation and mutual interference between radar and communication systems are studied for the coexistence of radar and multiple communication users. The main contributions of this paper are as follows:
\begin{itemize}
\item {The multicarrier model of~\cite{bica2016generalized} is extended to a more general setting in which multicarrier radar and communication systems coexist.  {In comparison to} ~\cite{bica2016generalized}, the resulting generalized multicarrier radar-communication signal model is applicable to a multicarrier radar-communication coexistence scenario and {thereby, shows} the differences between multicarrier radar and communication waveforms. }
\item {A joint precoder--decoder design is proposed using the max-SINR criterion and IA theory for a multicarrier-multiuser radar-communication coexistence scenario. {For receivers} {subject to} {cross-system interference,} the signal space and interference space are spanned by columns of the decoder. Consequently, mutual interference between radar and multiuser communication systems can be almost completely eliminated by the proposed joint~design.}
\item {For $K$ communication users and one radar user interference channel, with the assumption that the IA constraint is feasible, the proposed joint precoder--decoder design is able to achieve $N_{sc}(K+1)/2$ total DoFs, which is known as the achievable DoF upper bound for the $(K+1)$-user interference channel with $N_{sc}$ subcarriers~\cite{wu2011degrees}. In other words, if radar waveforms and communication codebooks are appropriately designed, this proposed IA-based design is able to achieve the optimal total information throughput for the entire radar-communication coexistence system. Radar {subsystem} could obtain better detection performance, diversity gain, and interference-free DoFs compared to a subspace-based precoder-only design.}
\end{itemize}

\subsection{Brief Overview of Related Work}
Various contributions have been presented in the radar and communication spectrum sharing literature. In general, these works can be classified into three main classes~\cite{chiriyath2017radar,bhattarai2016overview,li2022assisting,9540344}: codesign, cooperation and coexistence.
\begin{itemize}
\item \textbf{{Co-design}}: 
 {When hardware sharing between radar and communication systems} is possible, radar and communication systems can be jointly designed to maximize {their performances}~\cite{cui2022optimal,liu2021cram,liu2021integrated}. One example is a {Orthogonal Frequency Division Multiplexing (OFDM) dual-functional} radar-communication system: information transmission and target localization tasks can be 
{independently} and simultaneously accomplished by a {co-designed OFDM waveforms} \cite{sit2011ofdm,sit2011extension}. {Another example can be found in~\cite{blunt2010embedding, euziere2014dual, hassanien2015dual,hassanien2016signaling}, where communication information 
is} embedded into the sidelobe of the radar waveform to develop a co-designed dual-function system. 
\item \textbf{Cooperation}:  Limited information can be shared between radar and communication subsystems to {improve resource efficiency rather than isolate systems, e.g., to re-use shared spectrum by efficient interference managements}. {Bliss et al.} presented cooperative joint radar-communications inner bounds in~\cite{bliss2014cooperative,chiriyath2016inner,chiriyath2016jo} and extended this concept to MIMO systems~\cite{rong2017multiple}. Radar waveforms can be embedded as { a pilot signal for communication systems} in a doubly selective channel for detection and channel state estimation~\cite{harper2017performance}. {From our perspective, this work belongs to cooperative designs due to the limited information exchange, although we 
aim to suppress the interference via precoder-decoder co-designs.}
\item \textbf{Coexistence}:  {When radar and communication systems coexist and share spectrum in a non-cooperative mode, interference management becomes a key issue.} Practically, if the interfering energy is weak or the signal structure is unknown, an interfering signal may be treated as interference, e.g., interference from a Wi-Fi transmitter to a radar receiver. Furthermore, physical separation was introduced in~\cite{lackpour2011overview,hessar2016spectrum} to reduce the interfering energy below the noise level. Most of the prior radar-communication spectrum sharing approaches address interference management by exploiting {the orthogonality} property~\cite{wang2008application,saruthirathanaworakun2012opportunistic,bica2015opportunistic,sodagari2012projection,mahal2017spectral,khawar2014beampattern} or designing radar and communication signals while guaranteeing acceptable performance~\cite{li2016joint,li2016optimum,li2017joint,zheng2017joint}. A robust precoder that minimizes power {was proposed} for coexistence between MIMO radar and downlink multiuser MIMO communication systems in~\cite{liu2017interference}. The works in~\cite{chiriyath2016inner,chiriyath2017radar} explore information-theoretic  bounds for a single joint radar-communication user. The theoretical foundation of joint radar-communication research is established in these~studies.

\end{itemize}

 {However, the capacity of a multiuser radar-communication interference channel~\cite{carleial1975case,sato1981capacity,etkin2008gaussian} is still an open problem. An inspiring study~\cite{cadambe2008interference} showed that the IA scheme could achieve the total DoF upper bound as $\frac{K}{2}$ for a $K$-user interference channel. The key idea of IA is a linear precoding and decoding technique to design the signal space and orthogonal interference space for each user. {Essentially, the IA is a tradeoff between DoFs of signal space and interference space in a multiuser interference channel.}}

The definition of DoF is clear in the communication literature but is rarely studied in the {radar community}. In general, radar extracts information from targets~\cite{bekkerman2006target}. {Maximizing the DoFs upper bound corresponds to maximizing the achievable maximum number of interference-free measurements of the targets}. When the independent random variables follow an exponential distribution and the Neyman--Pearson detection strategy is applied, maximizing the frequency {DoFs corresponds to maximizing the diversity gain because the diversity gain for each independent variable in the exponential family is 1}~\cite{he2011diversity,mishra2022signal}. \textls[-25]{Although there is a large amount of results in the radar-communication coexistence category, the problem of interference management has not {been well investigated} in multiuser communication scenarios considering mutual interference and DoF exploitation simultaneously.}


\textbf{Notation}: In this paper, matrices are denoted by capital letters, vectors are denoted by boldface, $Re\{ \cdot \}$ means the real part of a complex signal, $\circ$ is the Hadamard product, $\mathbf{A}_{[i]} $ means matrix $\mathbf{A}$ for the $i$th transmitter and receiver, $\mathbf{A}_{[ij]} $ means matrix $\mathbf{A}$ with respect to the $i$th radio transmitter and $j$th radio receiver, $\mathbf{A}_{[R]} $  means matrix $\mathbf{A}$ with respect to the radar only, $\mathbf{A}_{N \times M}$ denotes an $N \times M$-dimensional matrix, $\lfloor \cdot \rfloor$ is the floor function, $\mathbf{A}^{H} $ is the Hermitian transpose of matrix $\mathbf{A}$, and $\overline{\mathbf{A}}$ is matrix $\mathbf{A}$ on the reciprocal interference channel. {Moreover, the math symbols are summarized in  {Abbreviations}. }

\section{Generalized Multicarrier Radar-Communication Coexistence Model}
In this section, we extend a generalized multicarrier radar signal model~\cite{bica2016generalized} to more general settings, where radar and communication systems  coexist and share the same radio spectrum.  {We start from a single-input single-output (SISO) point-to-point generalized multicarrier signal model to capture multicarrier radar and multicarrier communication signals.
Then, the signal model is applied for $K+1$ radar and communication users, where a radar user coexists with $K$ communication users.
}
\subsection{Transmitter}


 {\textls[-15]{The discrete-time signal model, {as detailed in Appendix A}, can be rewritten in a compact matrix~form as,}}
\begin{equation}\label{eqt} 
 \mathbf{Y_T} =  (\bm \Omega \circ \mathbf{B})   \mathbf{P}  \mathbf{C} \mathbf{S}
\end{equation}
where $\circ$ is the Hadamard product. $ \mathbf{Y_T}$ is an $ N_{sc} \times M$-dimensional sample matrix {at transmitter} that occupies $M$ time slots and $N_{sc}$ subcarriers for $k=\{0,1,2,...,MN_{sc}-1\}$ {transmit waveform}, denoted as a resource block for communication or pulse for radar. {The matrix $\bm \Omega$ is an $N_{sc} \times N_{sc}$-dimensional selection matrix, in which the elements of $\bm \Omega$ are either 1 or 0 to { activate or deactivate} subcarriers for either communication or radar subsystem. Matrix $\mathbf{B}\in \mathbb{C}^{N_{sc} \times N_{sc} }$ is considered to be a modulation matrix. The matrices $\mathbf{P} \in \mathbb{C}^{N_{sc} \times N}$  and $\mathbf{C}\in \mathbb{C}^{N\times N_p} $ respectively denote the linear frequency precoding matrix that contains  {subcarrier weights} and a coding
matrix, where $N_p$ denotes the length of the uncoded data sequence. Matrix $\mathbf{S}\in \mathbb{C}^{N_p\times M}$ contains payload data $\mathbf{S}= [ \mathbf{s}(1)... \mathbf{s}(M)]$, where $\mathbf{s}(M)$ is an $N_p$ data vector for $M$ time slots. } 

\begin{figure}[H]
	\centering
	\includegraphics[width=3in]{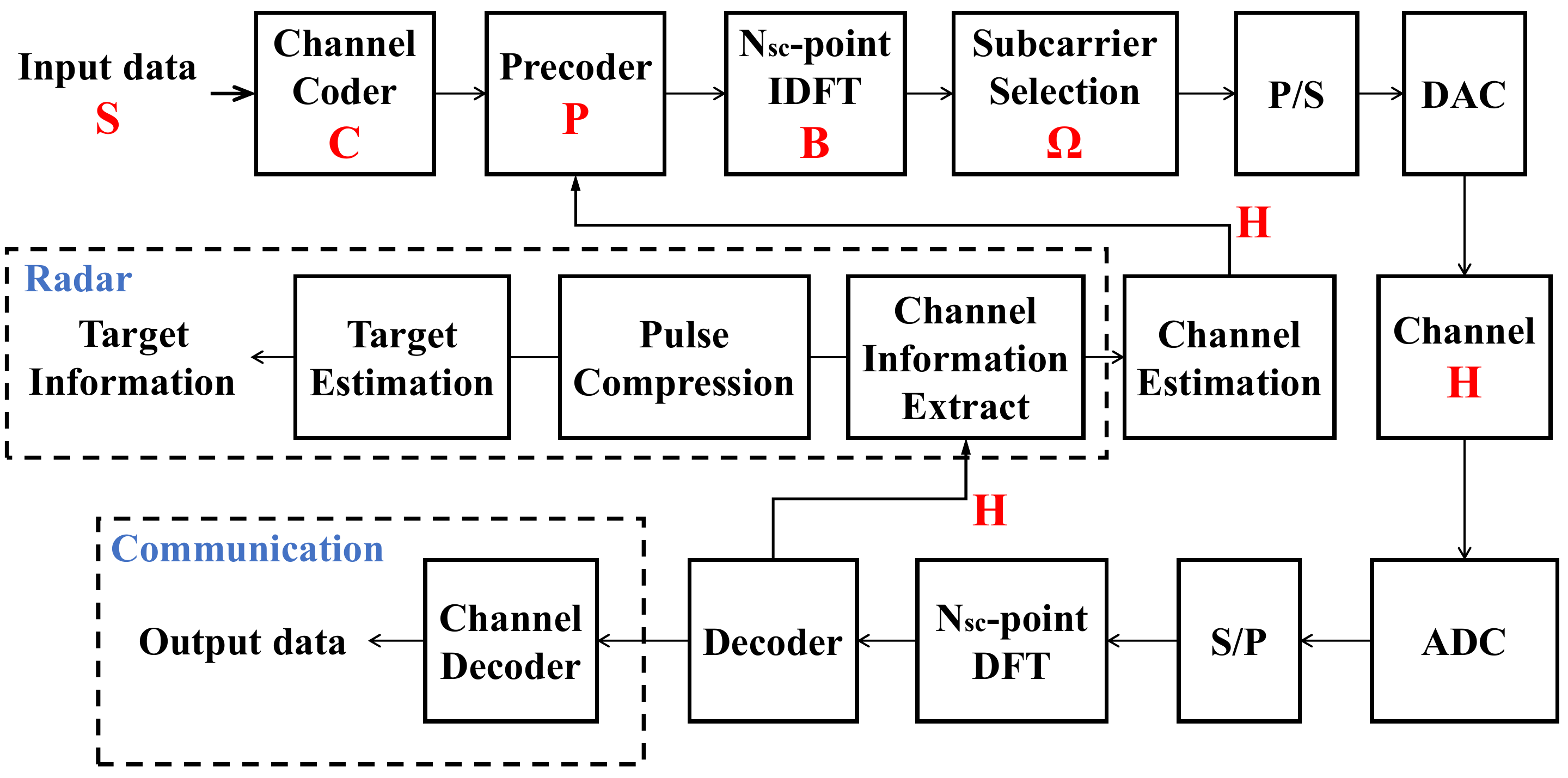}
\caption{{Multicarrier radio system for both multicarrier communication and multicarrier radar systems or a colocated radar-communication system. In the case of a radar or communication user, the dashed boxes of communication or radar in the block diagram are ignored. P/S (S/P) denotes parallel-to-serial or serial-to-parallel conversion, and DAC/ADC represents a digital-to-analog converter or analog-to-digital converter.}}
	\label{figure0}
\end{figure}

A block diagram of a multicarrier system for {co-designed radar and communication system} is shown in Fig.~\ref{figure0}.   This figure illustrates how the generalized multicarrier radar-communication signal model can be applied to generate signals at the transmitter and process them at the receiver. The input data will be processed by the channel coder, precoder, IDFT, subcarrier selection, parallel-to-serial {conversion and digital-to-analog converter to generate a multicarrier radar communication signal. After decoding, the channel information is estimated in the communication subsystem, or target information is estimated in the radar subsystem.} {In practice, the selection matrix selects carriers to construct a multicarrier waveform. Various waveform can be generated by 
enabling or disabling its elements, such as step approximation of up-chirp pulse, pseudo-random frequency-hopping pulse and multicarrier communication signal with a Comb-type pilot subcarrier. For example,
a widely used radar pulse and communication signal can be formed by deactivating resource block elements of $\bm \Omega$ \cite{bica2016generalized}.}

In an OFDM system, matrix $\mathbf{B}$ is a Vandermonde matrix, and its elements are associated with subcarriers as follows \cite{bica2016generalized}:
\begin{equation} \label{eqB}
\mathbf{B} =
 \begin{pmatrix}
   1 & 1 &\cdots  & 1 \\
   1 & \beta &\cdots  & \beta^{N_{sc}-1} \\
   1 & \beta^{2} &\cdots  & \beta^{2(N_{sc}-1)} \\
   \cdots & \cdots &  \cdots &  \cdots \\
   1 & \beta^{N_{sc}-1} &\cdots  & \beta^{(N_{sc}-1)(N_{sc}-1)}
  \end{pmatrix},
\end{equation}
{with each element {denoted by} $\beta^{(k \text{ mod } N)n}$
, where $n$ is the carrier index running of each columns and $\beta^{(k \text{ mod } N)n}$ denotes the sampling index of each row. Moreover, $\beta = e^{j2\pi \Delta f(\frac{T_c}{N_{sc}})}$
denotes baseband subcarriers without a carrier/sampling index, in which $T_c$ is the symbol duration of the multicarrier system, and $\Delta f$ is the subcarrier spacing.} The same active/inactive
subcarrier pattern is {applied} for the time duration of each radar subpulse/communication symbol. This pattern justifies the ($k$ mod $N_{sc}$) time duration above. In the special case of an OFDM signal,
matrix $\mathbf{B}$ is an inverse discrete Fourier transform (IDFT) matrix, $ \mathbf{B}^{H}\mathbf{B} =\mathbf{I}_{N_{sc}\times N_{sc}}$, where $T_c = \frac{1} {\Delta f}$.

In a multicarrier communication
system, the precoder functions as a multi-mode beamformer that allocates signal
power according to channel quality, similar to the MIMO spatial precoder in~\cite{vu2007mimo}. Furthermore, the power leakage could be {minimized} by an orthogonal precoding design~\cite{ma2011optimal}.  {For multicarrier radar,
$\mathbf{P}$ is applied as a multimode beamformer, projecting the radar signal into a designed subspace.} Moreover, the data channel coding rate is $\frac{N_p}{N}$ when coded data are transmitted in parallel and $N-N_p$ is the coding redundancy~\cite{declercq2014channel}. The code rate upper bound of coding matrix $\mathbf{C}$ is shown in~\cite{polyanskiy2010channel}. As channel coding increases the {transmission} reliability by introducing redundancy, it could increase the number of columns in the data matrix, i.e., $\frac{N_p}{N} \leq 1$. {Matrix $\mathbf{C}$ facilitates pulse compression
and consequently improves radar {range resolution}. However, it is indeed a channel coder in the communication subsystem to improve the reliability of communication transmission \cite{bica2016generalized}. In this paper, we employ the IA criterion to cancel the interference for both systems.}

As the transmitted waveform is always known {at the radar receiver}, without a loss of generality, we assume that the radar waveform is an all one matrix. Then, \mbox{$\mathbf{S} \mathbf{S}^{H} = M\mathbf{1}_{N_p\times N_p}$} for the radar signal. The transmitted payload data in the communication system is unknown to the receiver. It may be modeled with a complex Gaussian distribution, while its power is considered to be $\mathbb{E}(\mathbf{S}\mathbf{S}^{H})= \sigma_S^2 \mathbf{I}_{N_p\times N_p}$.

\subsection{Receiver}
Both the multicarrier radar and communication receiver in the {co-designed} system will obtain a radar response from the target reflection or an observation from an active transmitter in a communication system. For a radar receiver, assume that a single moving target
is located at direction $\theta$. By ignoring interference, a general form of the received signals before decoding  may be written as follows:
\begin{equation}\label{eq2}
      \mathbf{Y_{R}}
       = \mathbf{H}( \bm \theta ) (\bm \Omega\circ \mathbf{I}_{N_{sc} \times N_{sc}}) \mathbf{P}
       \mathbf{C}\mathbf{S}+\mathbf{W},
\end{equation}
where matrix $\mathbf{Y_R}$ is an $N_{sc} \times M$-dimensional received signal matrix. $\mathbf{W}$ denotes the \mbox{$N_{sc} \times M$}-dimensional complex white Gaussian noise that {obeys complex Gaussian distribution $\mathcal{CN}(\bm 0, \delta_W \mathbf{I}_{N_{sc} \times N_{sc}})$}. A detailed derivation of the model is presented in {Appendix \ref{app2}}.

The channel frequency response matrix denoted by $\mathbf{H}(\bm \theta)  \in \mathbb{C}^{N_{sc} \times N_{sc}}$ could be estimated for each $L$ symbol duration under the block fading channel assumption. {When an OFDM signal is employed, the channel for each subcarrier is assumed to be frequency flat and the channel state matrix is diagonal matrix $\mathbf{H}(\bm \theta)= \text{diag} \{ h_1(\theta), h_2(\theta) \cdots ,h_{N_{sc}}(\theta)\}$, where $h_1(\theta)$ is the channel frequency response at the first subcarrier }.

Considering the radar task at hand, the channel state matrix describes the combined effects of
target response, scattering, channel fading, Doppler shift, and power decay with distance. Among these parameters, the direction of arrival (DoA), {delay}, Doppler shift, and target response are of particular interest for the radar system. Note that if multi-antenna transceivers are used, spatial processing, such as DoA estimation may be performed. If a single-antenna system is employed, the DoAs can be estimated using a mechanically rotating directional antenna, as in many classical radar systems~\cite{skolnik1970radar}. The communication receiver treats Doppler shifts, multipath effects, and power decay as channel distortion. To ensure reliability of data transmission, channel coding and frequency precoding are typically employed. One needs to design a channel coding matrix $\mathbf{C}$ and a precoding matrix $\mathbf{P}$ to address the dynamic {nature of} ${\textbf{H}(\bm{\theta})} $.

\begin{figure}[H]
	\centering
	\includegraphics[width=3.5in]{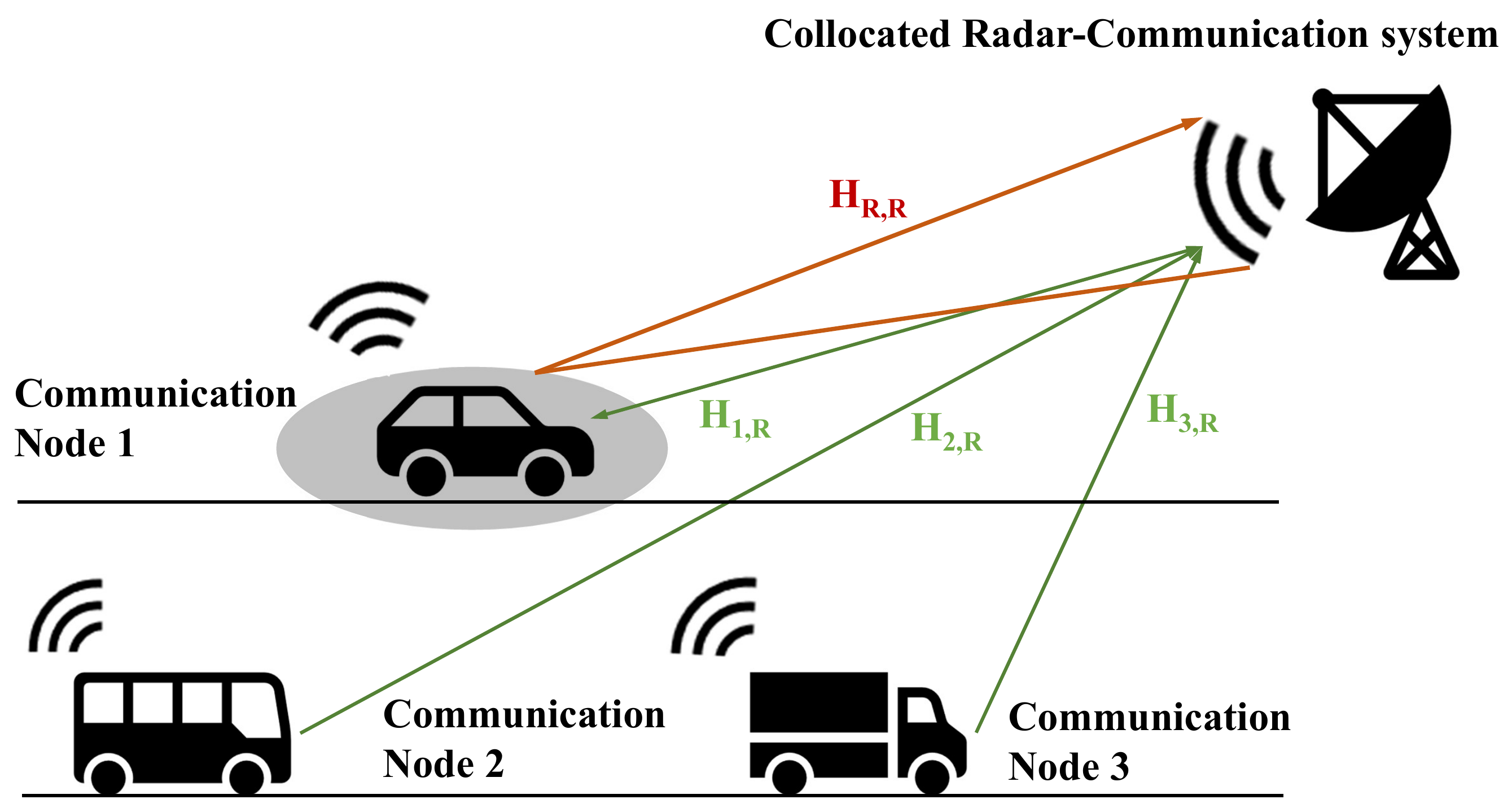}
	\caption{A practical vehicle networking example, in which the colocated radar-communication system coexists with a multiuser communication network}
	\label{figure2}
\end{figure}

In particular, for a co-located radar-communication node as shown in Figure~\ref{figure2}, the communication subsystem could easily share a channel state matrix with a radar subsystem. 
{Moreover, in cooperative settings if the target also carries a communication system, the performances of channel and target parameter estimation may both benefit from jointly estimating procedure}. A practical example of such a system is vehicle radar and vehicle-to-vehicle communication network.

\subsection{Multiuser Radar-Communication Spectrum Sharing Scenario}
In this  {subsection}, we consider a simplified $K+1$ user radar-communication spectrum sharing scenario, which consists of $K$ SISO communication users and a SISO multicarrier mono-static radar user. They share $N_{sc}$ subcarriers and the same frequency band.

Consequently, the {total number} of possible transmitter and receiver (TX-RX) pairs (including the signal channel and interference channel) is $K(K+1)$. {Each user is considered to be an intended TX-RX pair (matched link) transmitting a useful signal. Hence, the sum of useful signal channels is $K$. Interference occurred in unintended TX-RX pairs (mismatched link), which may be from radar TX to one communication RX, from one communication TX to radar RX or from communication TX to another unintended communication RX.} {In a nutshell,} the interference channels correspond to ${K^2}$ unintended links.

A mono-static radar is co-located with one of these communication nodes. {The co-designed radar communication system is considered to be a cooperating radar and a communication subsystem separately, as illustrated in Fig.~\ref{figure2}}. In this network configuration, we denote by $\mathcal{A}$ the set of communication users and by $\mathcal{B}$  the set of radar users, where
$\vert \mathcal{A} \vert = K, \vert \mathcal{B} \vert = 1 $. Here, $\vert \mathcal{A} \vert$ denotes the cardinality of set $\mathcal{A}$.
The co-located multicarrier mono-static radar illuminates one target in the direction $\theta$. {Note that radar is equipped with highly directional antenna} {or phased array} and is {capable of forming narrow beams}  such that it can only interfere with part of the communication nodes.

Recall that the signals are experiencing a block fading channel. Before decoding, the received multicarrier signal at the $i$th receiver in the $K+1$-user radar-communication coexistence scenario may be written as follows:
\vspace{12pt}
\begin{equation}\label{eq3}
 \begin{split}
\mathbf{Y_{R} }& =   \underbrace{\mathbf{H}_{[R]}(\theta)  (\bm \Omega_{[R]}
	 \circ \mathbf{I}_{N_{sc} \times N_{sc}} )\mathbf{P}_{[R]}\mathbf{C}_{[R]} \mathbf{S}_{[R]}}_{\text{Radar Signal}}\\
&+\underbrace{\sum_{j \in \mathcal{A}} \mathbf{H}_{[ij]} (\bm \Omega_{[j]}\circ \mathbf{I}_{N_{sc} \times N_{sc}})
\mathbf{P}_{[j]}\mathbf{C}_{[j]} \mathbf{S}_{[j]}}_{\text{Communication Signal}} \\
& +\mathbf{W}_{[i]}\\
\end{split}
\end{equation}
where $ \mathbf{Y}_{\mathbf{R}[i]}$ is {the $N_{sc} \times M$-dimensional} received signal for the $i$th user. Here, the $i$th receiver may refer to all $K$ communication receivers and radar receiver $i \in \mathcal{A} \cup \mathcal{B}$. $\bm \Omega_{[R]}$ denotes the selection matrix associated with {adaptive waveform design}, e.g., pseudo-random pulse or multi-pulse radar. Due to the block fading channel assumption, the channel coherence interval is $L$ pulses/resource blocks.  {Channel coding matrix $\mathbf{C}$ will remain unchanged because it is designed based on $\mathbf{H}$ in both communication and radar subsystems. Consequently, $\bm \Omega_{[R]}$, $\bm \Omega_{[j]}$, $\mathbf{P}$ and $\mathbf{S}_{[j]}$ may vary over $L$ pulses/resource blocks. }

To describe {this scenario} in detail, some key conditions are stated as follows:
\begin{itemize}
\item Channel State Information (CSI): The channels $\mathbf{H}_{[ii]}$ and $\mathbf{H}_{[ij]}$ and the target response {matrix} ${\textbf{H}(\bm{\theta})}_{[ii]}$ are considered to be perfectly estimated and {fed back to the communication and radar transmitters}, respectively.  {CSI estimation is commonly employed in communications systems. For radar, one feasible method is to treat a known radar waveform as a shared pilot in both radar and communication systems. } The pilot-aided approach can estimate all the channel information between radar and communication users~\cite{harper2017performance}. Another approach is to embed the same pilot signal in both the radar coherence interval and communication frames~\cite{li2017joint}. 
Moreover, channel reciprocity may be exploited for interference channels.
\item Synchronization: Both radar and communication systems are assumed to be synchronized. If they are colocated, then they may share the same clock, in which case, synchronization is not an issue. The other subsystems need to be synchronized in a similar manner to any multiuser communication system; the clock synchronization may be easier in communication systems but still feasible for radar. Existing radar clock synchronization technology may be employed, such as using a Global Navigation Satellite System (GNSS)~\cite{yulin2006synchronization,wang2009gps}, using a pilot signal~\cite{wang2007approach} or using an OFDM frame~\cite{schmidl1997robust} to achieve time and frequency synchronization~\cite{sit2011ofdm}.
\item Shared Information: As discussed above, the following information is shared among all users:  {selection matrix $\bm \Omega^{(l)}$ and communication power $\sigma_S^2 \mathbf{I}_{N_p\times N_p}$}. This shared information {will be employed} to calculate the transmitted signal power and consequently to solve the following optimization problem {in the next section}. {Alternatively, the transmitted signal power can be constrained by the plimited by a power constraint.}
\item Doppler and schedule: The Doppler shift is assumed to be constant during a coherent interval for the $L$ pulse. The feedback of the channel state matrix, transmission of clock synchronization and shared information call for a  {protocol for exchanging information between the communication and radar subsystems}. Providing channel feedback is common in most modern communication systems and part of the standards. A radar system can also take advantage of feedback and estimate channels. One feasible approach is to transmit information between the radar coherent interval ~\cite{li2017joint}.
\end{itemize}

\section{Max-SINR Joint Precoder--Decoder Design}

{IA} is an emerging DoF-based interference management technique in wireless communications that aligns the interference caused by other users in an interference signal subspace that is orthogonal to
the user-desired signal subspace~\cite{gomadam2011distributed}. This technique can be applied in the time, frequency or spatial {domains}. Furthermore, in a high-SNR regime, this technique can achieve high-interference elimination and offer a $\frac{K}{2}$ achievable total DoF upper bound in the $K$-interference-communication-user scenario without frequency/time/spatial symbol extension~\cite{cadambe2008interference,wu2011degrees}. Symbol extension means additional diversity, for example, by expanding bandwidth or adding antennas in this paper.
At the same time, the traditional orthogonal spectrum allocation, which allocates a nonoverlapping signal to each user, could only obtain $\frac{1}{K}$ interference-free DoFs.

In this section, we propose a joint precoder--decoder design using the max-SINR criterion and IA approach to solve the radar-communication {spectrum sharing} problem.
 {Communication receivers are interfered with by radar via a direct path.} As the radar signal reflected from the target may also have a high power,  {we could treat the scattered radar signal as noise for the communication receiver and consequently a medium SNR scenario}. IA in a high-SNR regime typically focuses on
minimizing the leakage interference while ignoring noise. However, in this radar-communication coexistence problem in a medium-SNR regime, it is necessary to take the noise into account. Hence, we will employ the max-SINR criterion for our design.

\subsection{Ideal IA Constraints}
{The ideal IA constraints in a multicarrier radar-communication coexistence scenario can be written as {follows}~\cite{gomadam2011distributed}}:
\begin{subequations}\label{IA1}
 	     \begin{gather}\label{IA11}
      \mathbf {Q}_{[i]}^H \mathbf{H}_{[ij]} \mathbf{P}_{[j]}  = \mathbf{0}_{{d_{[i]} \times d_{[j]}}} ,  \\
     \text{rank}( \mathbf{Q}_{[i]}^H \mathbf{H}_{[ii]} \mathbf{P}_{[i]} ) = d_{[i]} , \label{IA12}\\
    \forall i \neq j ;  \forall i,j \in \mathcal{A} \cup \mathcal{B} ,\label{IA14}
   \end{gather}
 \end{subequations}
where $ \mathbf{0}_{{d_{[i]} \times d_{[j]}}}$ is a $d_{[i]} \times d_{[j]}$ zero matrix, $\mathbf{Q}_{[i]}$ is an $N_{sc} \times N_{[i]}$-dimensional decoder matrix of the $i$th communication node, and recall that $N_{[i]}=d_{[i]}$.
 $d_{[i]}$ denotes the {number of} user-desired DoFs for the $i$th
user. The channel matrix of the interference channel between the $i$th radio node and the $j$th radio node is $\mathbf{H}_{[ij]}$, {which is known to both radar and communication subsystems}, and the matrix of the signal channel at the $i$th user is $\mathbf{H}_{[ii]}$. When $i \in \mathcal{A}$, $\mathbf{H}_{[ii]}$ is a target response matrix $\mathbf{H}_{[ii]}(\theta)$, which is of interest to the radar subsystem.

{Constraint \eqref{IA11} means that the precoder is designed to project the interference from the $j$th radio transmitter to the $i$th receiver's nullspace.} The nullspace is designed by decoder $\mathbf{Q}_{[i]}$ to align interference from all $j$ transmitters, $  \forall i \neq j ; \forall j \in \mathcal{A} \cup \mathcal{B}$. Equation \eqref{IA12} implies that the designed signal space should {provide the number of desired DoFs}. The ideal IA constraints design precoders $\mathbf{P}_{[j]}$ to project interference such that it is aligned in the nullspace of $\mathbf{Q}_{[i]}$ and design $\mathbf{Q}_{[i]}$ to guarantee that the signal space and interference space exits. {Noting \eqref{IA12} must be taken into account due to the diagonal channel state matrix of the OFDM channel, though constraint \eqref{IA12} is automatically satisfied almost surely in the MIMO configurations such that it can be ignored.}

 \subsection{Reciprocity and Feasibility of IA}
{In a time division duplex (TDD) radio link}, the roles of the receiving and transmitting antennas are functionally interchanged, while the instantaneous transfer characteristics of the radio channel remain unchanged.
This channel reciprocity can be exploited in IA design. The reciprocity of IA~\cite{gomadam2011distributed} comes from the identity IA constraints between the original interference channel and the reciprocal interference channel, where the original precoder and decoder are considered to be a reciprocal decoder and  precoder, respectively. In other words, IA constraints are still feasible after the signal direction is reversed. Using reciprocity, the IA constraints of {\eqref{IA1}} can be written {as follows:}
  \begin{equation} \label{IAintuitiveReceprocal}
 	\mathbf{0} =  \overline{\mathbf {Q}}_{[j]}^H \overline{\mathbf{H}}_{[ji]} \overline{\mathbf{P}}_{[i]},
 	\mathbf{\overline{\widetilde{H}}}_{[ii]} =  \overline{\mathbf {Q}}_{[i]}^H \overline{\mathbf{H}}_{[ii]} \overline{\mathbf{P}}_{[i]}, \\
 	\forall i \neq j ;  \forall i,j \in \mathcal{A} \cup \mathcal{B}
 \end{equation}
\textls[-35]{{where $\overline{\mathbf{P}}$, $\overline{\mathbf {Q}}$ and $\overline{\mathbf{H}}_{[ii]}$ denote the precoder, decoder and channel matrix on the reciprocal channel, respectively, where} $\overline{\mathbf {Q}}_{[i]}=\mathbf {P}_{[i]}, \overline{\mathbf {P}}_{[i]}=\mathbf {Q}_{[i]},$} $\forall i \in \mathcal{A} \cup \mathcal{B}$. Furthermore, $\mathbf{{0}}_{{d_{[i]} \times d_{[j]}},[ij]}$ denotes the interference nullspace from the $i$th receiver to the $j$th transmitter on the reciprocal channel. $\mathbf{\overline{\widetilde{H}}}_{[ii]}$ denotes the channel matrix for the $i$th user itself on the reciprocal channel. The reciprocity of IA does not change the user-desired DoFs of each user while projecting an undesired signal into the nullspace. However, it plays an important role in the following distributed {iterative} algorithm {proposed in this paper}.

As with the communication signal, the DoFs of a radar subsystem need to be properly chosen to realize IA. The radar subsystem also {desires more available DoFs that facilitates more flexible waveform design and a higher diversity gain}. In this paper, we assume that the number of desired DoFs is predetermined. The DoFs of radar need to satisfy the feasibility condition of IA, which is written as follows:
\begin{equation}\label{IA13}
  \begin{cases}
  &d_{[R]}+d_{[i]} \leq N_{sc} ,\forall i\in \mathcal{A} \cup \mathcal{B} \\
  &2d_{[R]}(N_{sc}-d_{[R]})-\sum_{i\in \mathcal{A}_{sub}}d_{[R]}d_{[i]} \geq 0,\forall \mathcal{A}_{sub}\subset \mathcal{A} \cup \mathcal{B}. \\
  \end{cases}
 \end{equation}
 
 If IA is feasible, then the dimension of projection between the strategy space and channel matrix must be non-negative~\cite{bresler2011feasibility}. This condition can easily be formulated based on Theorem 2 in~\cite{bresler2011feasibility}. Note that \eqref{IA13} is a necessary condition. IA requires finding feasible signal strategies while the channel matrix is fixed. However, the feasibility is established in a reverse way, where the strategy is fixed and the study channel matrix is set for which strategy is feasible. The space of strategies can be represented by the product of Grassmannians ($7$, \cite{bresler2011feasibility}). Furthermore, if $d_{[i]}=d_{[R]}=d$, i.e., the desired DoFs of the communication and radar nodes are identical and equal to $d$, the IA is feasible if and only if
\begin{equation}\label{IAfe}
d \leq \frac{2N_{sc}}{K+1}.
 \end{equation}
 
This condition degenerates into a necessary condition when $d_{[i]}=d_{[R]}=d=1, K\geq 3$, where {each users call for a single data stream} (Theorem 1,~\cite{du2014feasibility}) 
 \begin{equation}
 K \leq 2N_{sc}-2.
 \end{equation} 
  
 {This solution is a degenerated}  {solution} {of \eqref{IAfe}, i.e. $2N_{sc}-1$, because there will be at least one remaining channel not used by any transmitter for each IA solution in multicarrier scenario.} The general feasibility condition of a proper IA network with multiple streams for each user remains an open problem.

 \subsection{Distributed {Max-SINR} Precoder--Decoder Design}

 Based on IA theory, \eqref{IA11}--\eqref{IA14} could then be formulated as an  {interference minimization problem  {as follows:}} 
    \begin{equation}
 	\begin{aligned}
 	\underset{\mathbf{P}_{[i]},\mathbf{Q}_{[i]}}{\min} &Tr(\mathbf{Q}_{[i]}^{H} \mathbf{H}_{[ij]}
 	\mathbf{P}_{[j]}) \label{1a}	\\
 	s.t. \quad
 	&\text{rank}( \mathbf{Q}_{[i]}^{H} \mathbf{H}_{[ii]} \mathbf{P}_{[i]}) = d_{[i]}. 
 	\end{aligned}
 \end{equation}
 
 {However, the interference minimization criteria are not optimal when taking noise and the radar channel matrix into account.}
 {Without loss of generality, \eqref{IA11}--\eqref{IA14} could be formulated as a SINR optimization problem {as follows:} } 
\begin{subequations}\label{IAoptimal}
   	   \begin{alignat}{1}
       \underset{\mathbf{P}_{[i]},\mathbf{Q}_{[i]}}{\max} &Tr(\frac{\mathbf{Q}_{[i]}^{H} \mathbf{H}_{[ii]}
       \mathbf{A}_{[i]}\mathbf{A}_{[i]}^{H} \mathbf{H}_{[ii]}^H
        \mathbf{Q}_{[i]} }{
       \mathbf{Q}_{[i]}^{H} (\sum_{j=1}^{Z_{[i]}} \mathbf{H}_{[ij]}  \mathbf{A}_{[j]}\mathbf{A}_{[j]}^{H} \mathbf{H}_{[ij]}^H+ \delta_W \mathbf{I}
        ) \mathbf{Q}_{[i]} })\label{IAo1}\\
       s.t. \quad
       &\text{rank}( \mathbf{Q}_{[i]}^{H} \mathbf{H}_{[ii]} \mathbf{P}_{[i]}) = d_{[i]}, \label{IAo2} 
     \end{alignat}
   \end{subequations}
{where $Z_{[i]}$ is the number of interfering sources for the $i$th receiver.} Projecting interference into the designed nullspace with fixed SNR implies maximizing the SINR as \eqref{IAo1}. Let us recall \eqref{eqt};
the signal power emitted from the $i$th transmitter may be written as
\begin{equation}\label{calculateA}
 \mathbf{A}_{[j]} \mathbf{A}_{[j]}^{H}=
\begin{cases}
M (\bm \Omega_{[R]}\circ \mathbf{I}_{N_{sc} \times N_{sc} })
\mathbf{P}_{[R]} Tr(\mathbf{C}_{[R]} \mathbf{1} \mathbf{C}_{[R]}^{H}) \mathbf{P}_{[R]}^{H}  (\bm \Omega_{[R]} \circ \mathbf{I}_{N_{sc} \times N_{sc} })^H
\text{, if  } j \in \mathcal{B} \\
\sigma_{S}^2(\bm \Omega_{[j]}\circ \mathbf{I}_{N_{sc} \times N_{sc} })
\mathbf{P}_{[j]} Tr(\mathbf{C}_{[j]}  \mathbf{C}_{[j]}^{H}) \mathbf{P}_{[j]}^{H}  (\bm \Omega_{[j]} \circ \mathbf{I}_{N_{sc} \times N_{sc} })^H
\text{, if } j \in \mathcal{A}.
\end{cases}
\end{equation}
{where $\mathbf{1}$ is a $N_p \times N_p$-dimensional all-ones matrix.} Recall that the set  $\mathcal{A}$ contains communication users. Some of these communication receivers are interfered by radar, denoted as the subset $\mathcal{A}_r$, and the complement set of communication users not interfered by radar is denoted as $\mathcal{A}_c$. {If the $i$th communication receiver is interfered by radar, $i \in \mathcal{A}_r$, it will experience interference from $Z_{[i]} = K-1$ communication transmitters
and one radar transmitter. 
If this receiver is not interfered by radar, $i \in \mathcal{A}_c$, its interference
is caused by $Z_{[i]} = K-1$ communication transmitters only.
If this receiver is a radar receiver, then it is subject to interference from all $Z_{[i]} =K$ communication transmitters.} 
Here, we use the commutative law of the Hadamard product and the generalized radar-communication coexistence signal model in Section \uppercase\expandafter{\romannumeral2}-C.
The trace term in \eqref{calculateA} corresponds to the signal power before the precoding and modulation operation.

The objective function is maximized in an iterative manner. In each iteration, we maximize the objective function  \eqref{IAo1} to find  $\mathbf{Q}_{[i]}$  at each  receiver. Then, at each transmitter, we will find the original $\mathbf{P}_{[i]}$ according to  \eqref{IAo1}. At this time, \eqref{IAo1} can be further simplified to accelerate the calculations.

The interference plus noise covariance matrix for the $i$th receiver may be written {as~follows}:
\begin{equation}\label{calculateB}
\mathbf{D}_{[i]} =\sum_{j=1}^{Z_{[i]}}\mathbf{H}_{[ij]} \mathbf{A}_{[j]} \mathbf{A}_{[j]}^{H} \mathbf{H}_{[ij]}^H+ \delta_W \mathbf{I}.
\end{equation}

Recall that all terms except $\mathbf{Q}_{[i]}$ are fixed to find $\mathbf{P}_{[i]}$. When maximizing \eqref{IAoptimal}, $\mathbf{Q}_{[i]}$ is normalized to limit the decoder element scope for eigenvectors and have easier calculations. It may be given by the following:\vspace{12pt}
\begin{equation}\label{calculateQ}
\mathbf{Q}_{[i]} = \frac{\mathcal{V}_{d_{[i]}}( \mathbf{D}_{[i]}^{-1}\mathbf{H}_{[ii]} \mathbf{P}_{[i]}\mathbf{P}_{[i]}^{H}\mathbf{H}_{[ii]}^{H})}{\|\mathcal{V}_{d_{[i]}}( \mathbf{D}_{[i]}^{-1}\mathbf{H}_{[ii]} \mathbf{P}_{[i]}\mathbf{P}_{[i]}^{H}\mathbf{H}_{[ii]}^{H})\| },
\end{equation}
where $\mathcal{V}_{d_{[i]}}(\mathbf{A})$ denotes the eigenvectors corresponding to the $d_{[i]}$ smallest eigenvalues of $\mathbf{A}$. 

The objective function in \eqref{IAoptimal} is not convex. However, we could find a solution by {exploiting channel reciprocity and employing a distributed iterative algorithm}. This algorithm is guaranteed to converge, but it may not necessarily find the global optimum~\cite{gomadam2011distributed}. By taking all radio nodes into account, we can write an algorithm for solving this optimization problem in \eqref{IAoptimal}.
See Algorithm~\ref{alg:A} for detailed steps. The reciprocal interference channel is still IA feasible by choosing the original precoder as a decoder  of the reciprocal interference channel and the original decoder as a precoder, respectively. The algorithm finds $\mathbf{Q}_{[i]}$ and $\mathbf{P}_{[i]}$ iteratively in two stages.

\begin{algorithm}
\caption{Max-SINR design algorithm.}
\label{alg:A}
\begin{algorithmic}[1]
\STATE {Estimate radar channel $\mathbf{H}_{[R]}$ and radar's interference channel $\mathbf{H}_{[iR]}$ }
\STATE {Initialize $\mathbf{P}_{[i]}$,$\mathbf{Q}_{[i]}$ with independent row vectors, $\forall i\in \mathcal{A} \cup \mathcal{B}$}
\REPEAT
\REPEAT
\STATE Identify location and type of $i$th node,  $\forall i\in \mathcal{A} \cup \mathcal{B} $
\STATE Choose $Z_{[i]}$ according to its location and type
\STATE Calculate transmitted signal power according to equation \eqref{calculateA}
\STATE Calculate interference pulse noise covariance matrix $\mathbf{D}_{[i]}$ according to equation \eqref{calculateB}
\STATE Find  $N_{sc} \times d_{[i]} $ matrix $\mathbf{Q}_{[i]}$  on each receiver according to \eqref{calculateQ}
\UNTIL{All $\mathbf{Q}_{[i]}$ are found}
\REPEAT
\STATE Use channel reciprocity
\STATE Identify location and type of $i$th node,  $\forall i\in\mathcal{A} \cup \mathcal{B} $
\STATE Choose $Z_{[i]}$ according to its location and type
\STATE Calculate power of reciprocal transmitted signal  according to equation \eqref{calculateA}
\STATE Calculate  interference pulse noise covariance matrix $\mathbf{D}_{[i]}$ of reciprocal signal according to equation \eqref{calculateB}
\STATE Calculate    $N_{sc} \times d_{[i]} $ matrix $\mathbf{P}_{[i]}$  on each reciprocal receiver according to \eqref{calculateQ}
\UNTIL{All $\mathbf{P}_{[i]}$ are calculated}
\STATE Check rank of $\mathbf{Q}_{[i]}^{H} \mathbf{H}_{[ii]} \mathbf{P}_{[i]}$ to verify (14b).
\UNTIL (14b) is satisfied and {\eqref{IAoptimal} has converged or {maximum allowed iteration count is achieved.}}
\end{algorithmic}
\end{algorithm}

The proposed distributed algorithm operates for finding precoders and decoders in an alternating manner. We start by
finding solutions for $\mathbf{Q}_{[i]}$ at each receiver from \eqref{IAoptimal}, with fixed $\mathbf{P}_{[i]}$, $\mathbf{P}_{[j]}$, $\mathbf{Q}_{[j]}$,$	\forall i \neq j ;  \forall i,j \in \mathcal{A} \cup \mathcal{B}$. Precoders $\mathbf{P}_{[i]}$ are found using fixed decoders $\mathbf{Q}_{[i]}$, $\mathbf{Q}_{[j]}$, $\mathbf{P}_{[j]}$, $	\forall i \neq j ;  \forall i \in \mathcal{A} \cup \mathcal{B}$, where $j$ is selected based on its node type and location.
In the next stage, we will reverse the signal direction, the original transmitter will be treated as a receiver, and the original receiver will be treated as a transmitter. {Consequently, decoders of reciprocal channel be found, which are indeed $\mathbf{P}_{[i]}$ of the original channel in \eqref{IAoptimal}, by fixing the other terms $\mathbf{Q}_{[i]}$, $\mathbf{Q}_{[j]}$ and $\mathbf{P}_{[j]}$.}  For each precoder and decoder, the rank of signal space is checked after one solution is obtained. This iteration will continue until convergence, which is evaluated by comparing \eqref{IAo1} to a threshold value or until the iteration count reaches its maximum value.

The constraint \eqref{IAo2} may also be relaxed by considering matrix dimension $N$ as $d_{[i]}$ at the $i$th user. This condition guarantees that a trivial and useless solution of all zeros is avoided.
Consequently, we  design an $N_{sc} \times d_{[i]}$ precoder and a $ d_{[i]} \times N_{sc} $ decoder for the $i$th node.

{Note that the precoder and the decoder for both radar and communication subsystems in a colocated node could be more conveniently found by the proposed algorithms. In cases where radar and communication subsystems are colocated and consequently suffer from the same interference, the interference channel matrix estimated by the communication system could easily be re-used for radar precoder and decoder designs. } Furthermore, if noise statistics are the same in both radar and communication subsystems when sharing the same hardware architecture, the interference plus noise covariance matrix is also identical. {It could be shared between the subsystems to reduce calculation time, by using the same memory or optical fibers.}

\section{Simulation Examples}

In this section, we present simulation results to demonstrate the performance of the proposed max-SINR
joint precoder--decoder design algorithm. {The simulation is done using MATLAB 2019b and a desktop with an i7-10700k processor.  We consider a four-user interference channel where three communication users and one radar user colocated with one of the communication users.} Each node employs a multicarrier signal model with $N_{sc}=8$ subcarriers. {Assume that communication users desire one DoF and
radar  desires three DoFs, which is the highest practical achievable DoFs for the radar subsystem under~\eqref{IA13}.}  {We also assume that zero mean Gaussian noise are independent and with variance $\sigma_{W}^2=1$ is present at each receiver. The power of the payload signal is $\sigma_S^2 =1$ for each communication user. Consider orthogonal channel coding is employed such that coding matrix} $\mathbf{C}_{[i]}\mathbf{C}_{[i]}^H = \mathbf{I}$  is an identity matrix. The far-field point target is at azimuth angle $\theta = 0^\circ$. The total number of samples is 500. The initial estimates of the precoder and decoder are obtained from independent row vectors. One of these three communication users is not interfered with by radar.
Increasing the transmit power at each transmitter will also increase the interference observed by the other receivers. {The channel matrices are generated randomly according to the block fading channel assumption, where each entry of the channel matrix follows the standard complex Gaussian distribution. Moreover, we assume that the entries of the target response matrix follows Swerling 2 model with Gaussian distributed complex amplitude}.

We evaluate the system performance by
the sum of all the SINRs in the coexistence scenario. {The performances of the communication and radar subsystems clearly depend heavily on the SINR values at the receivers. Radar performance in the detection task is studied using a ROC curve, which evaluates the performance of a radar detector by plotting the probability of detection (Pd) versus the probability of false alarm (Pfa) for a given conditions}.

We compare the performance of the proposed method with the switched small singular value space projection (SSSVSP) in~\cite{mahal2017spectral}. SSSVSP is an extension of nullspace-based precoder design in which the nullspace has been expanded to include the subspace spanned by singular vectors. These singular vectors correspond to small singular values that are selected based on a threshold value. However, it still faces the drawback that it cannot handle mutual interference between radar and communication systems. In the following simulations, we compare the proposed algorithm with SSSVSP when interference from radar is projected into communication users' switched small singular value space.

\begin{figure}[H]
\centering
\includegraphics[width=4.2in]{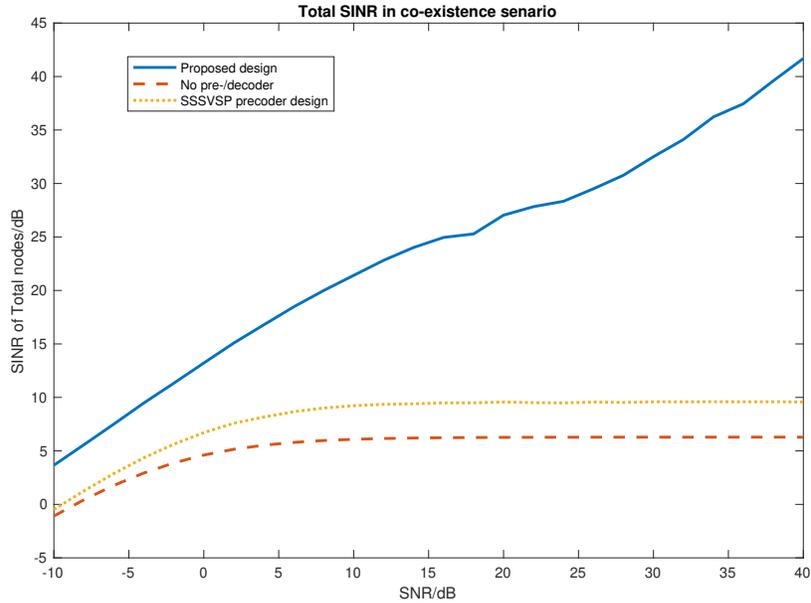}
\caption{{{The total} SINR, which is the sum of SINR values at all receivers for the {proposed design, SSSVSP}, and original signal for coexistence between the radar and communication~network.} }
\label{figure3}
\end{figure}

In Figure~\ref{figure3}, we compare the total SINR values from all radio receivers, including communication and radar. For fairness in the comparison of DoFs,  the number of {transmit and receive antennas} in SSSVSP is selected to be eight.  {Noting the multicarrier interference channel matrix is in general a diagonal matrix.}
For the proposed method, the SINR increases as a function of SNR, whereas the other approaches experience more interference, and consequently their SINR values decrease. However,
the SSSVSP and {a design without precoder or decoder method} decrease with it. 

This result may be due to the projection of the radar signal to the nullspace of all communication receivers while ignoring the interference experienced at the radar receiver and interferences among the communication nodes. Moreover, if the signal powers increase at the transmitters, it will increase the interference power at the other receivers. {The SINR of the proposed design increases as a function of SNR because interferences are almost completely eliminated.}
We also observe a good total SINR improvement in a medium-SNR regime.

\begin{figure}[H]
\centering
\includegraphics[width=\linewidth]{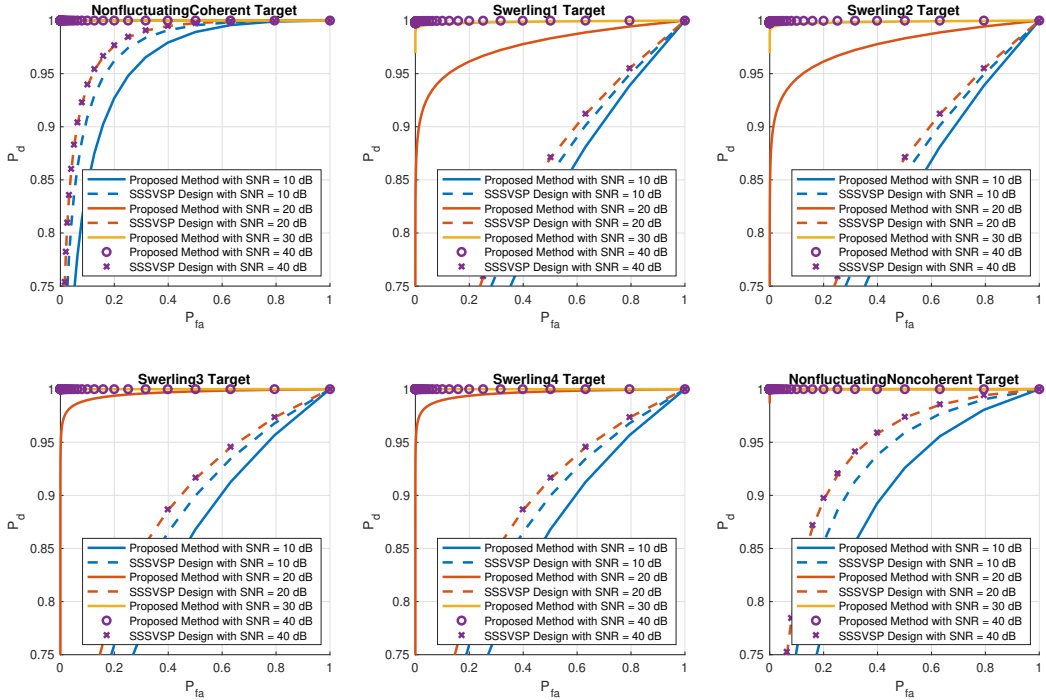}
\caption{{The radar's} {ROC curve comparison between the proposed max-SINR joint precoder--decoder design and SSSVSP design in 0, 10, 20 and 30 dB SNR. Comparison results under various target models are illustrated in above figures. The proposed method provides superior performance in all considered scenarios.}}
\label{figure4}
\end{figure}

{The detection performance of proposed design and SSSVSP method are compared using Neyman-Pearson detectors under different false alarm constraints, their ROC curves are plotted in Fig. \ref{figure4}}. Different SNR levels are considered. {In this simulation, one radar and one communication user are coexisting, where nonfluctuating and coherent target model, Swerling I-IV target, and nonfluctuating but coherent target model are investigated in the simulation, respectively.} All radio transmitters are assumed to be active such that every receiver suffers from interference from other transmitter. {The probability of detection as a function of SNR curves is drawn by using 500 radar pulses}.

{Figure~\ref{figure4} shows that the proposed max-SINR joint precoder--decoder design outperforms than the SSSVSP design when a radar coexists with one communication user in medium and high SNR regions, namely SNR = 20, 30, and 40 dB. When the SNR is relatively low, i.e. below SNR= 10 dB, the interference plus noise matrix is dominated by the noise power. In this case, the SSSVSP design is employed at communication side, and therefore projects the communication interference into the alternative signal space so that the radar performance is slightly better than our proposed method. Moreover, one can also observe that the detection performance does not increase with the SNR by comparing the curves of SSSVSP design with SNR = 20 dB and SNR = 40 dB. Therefore, there always be some interference leakage with the SSSVSP strategy.}

\begin{figure}[H]
	\centering
	\includegraphics[width=4.5in]{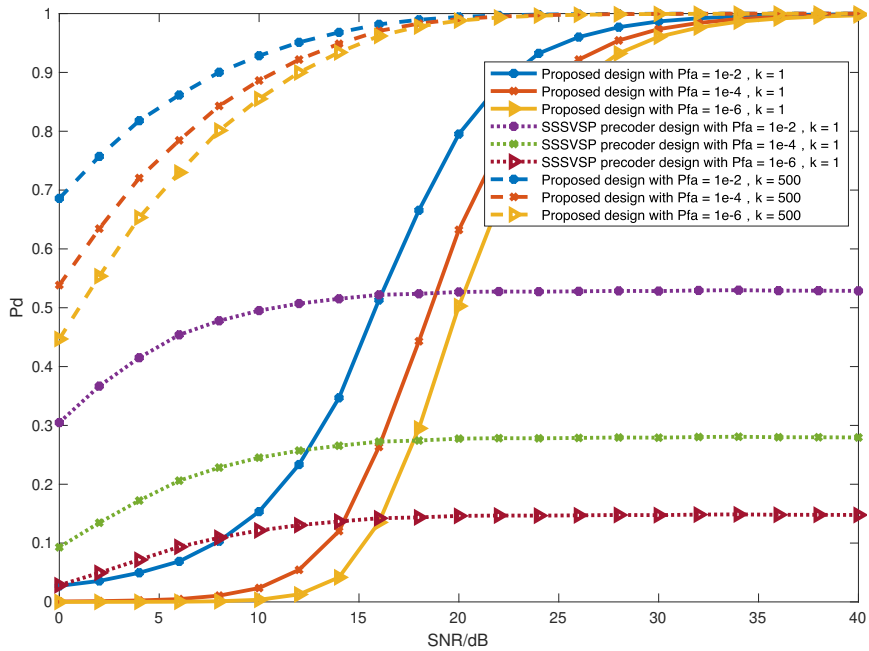}
\caption{Detection performance difference comparison between proposed max-SINR joint precoder--decoder design and SSSVSP at $P_{fa} = 10^{-2},P_{fa} = 10^{-4}$, and $P_{fa} = 10^{-6}$. A single pulse $k=1$ and multi-pulse $k=500$ are considered in this comparison.}
	\label{figure5}
\end{figure}

{In Figure~\ref{figure5}, the difference in radar performance between the proposed max-SNR joint precoder--decoder design and SSSVSP is shown. A single pulse $k=1$, i.e. without pulse compression, and multi-pulse $k=500$, i.e. with pulse compression, are considered. In this simulation, $N_{sc}$ are set to be 16 and the target model is Swerling I. A Neyman--Pearson detector is applied with false alarm constraints $P_{fa} = 10^{-2},P_{fa} = 10^{-4}$, and $P_{fa} = 10^{-6}$. It can be observed that SSSVSP design achieves the detection performance upper bound at a relative low SNR region without pulse compression technology. By increasing the number of pulses, the radar detection performance improves because the radar performance  benefits from coherent signal processing.}

\begin{figure}[H]
	\centering
	\includegraphics[width=5in]{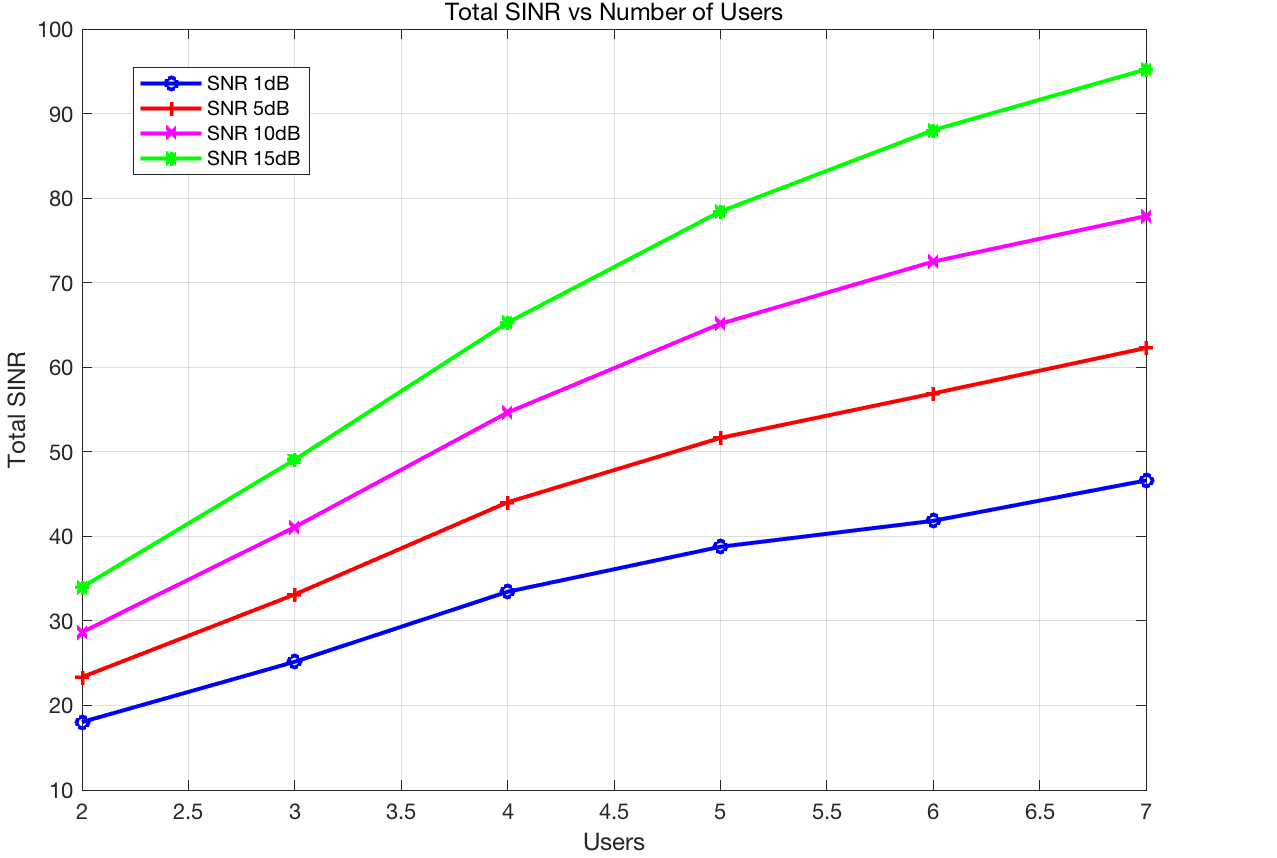}
\caption{{System} {total sum SINR performance} as a function of the number of users for the proposed design. The user-desired DoF is 1.}
	\label{figure7}
\end{figure}

In Figure~\ref{figure7}, the SINR performance is studied as a function of the number of users. In this simulation, the DoF of each user is 1. The results show that total system SINR increases as the number of users increases. An additional user may add a useful signal to the system. However, it will cause interference for the other users. The proposed design can successfully remove interference for the entire system. The higher SNR at each receiver could also increase the system SINR. The proposed design avoids performance loss when the number of users increases.

\section{Conclusions}
{This paper considered the problem of interference management and alignment in radar and communication spectrum sharing scenarios. A generalized multicarrier signal model is {employed}. The model provides an easy way {to represent}, generate and analyze a multicarrier signal. }A max-SINR precoder-decoder joint design based on IA theory is proposed. The design is formulated as a constrained optimization problem. By employing IA theory, the benefit of achieving the total DoF upper bound can be obtained. {A distributed and alternating algorithm for finding the precoder and decoder is derived as a solution. It takes advantage of channel reciprocity in TDD system.} The proposed design allows for achieving {better interference suppression }between communication and radar nodes in comparison {to existing} precoder design.
The simulation results demonstrated that our algorithm significantly improves the total SINR for
all radar and communication nodes and provides a significantly higher target detection probability in a radar subsystem at a given constraint {using Neyman-Pearson detector while making the communication system almost interference-free}. The sum SINR increases as a function of SNR and the number of users while avoiding performance degradation.

\authorcontributions{{Conceptualization, Y.C. and V.K.; Formal analysis, Y.C.; Funding acquisition, X.J.; Methodology,  Y.C. and V.K.; Project administration,  Y.C. and X.J.; Resources, Y.C.; Software, Y.C.; Writing---review \& editing, Y.C. and V.K. All authors have read and agreed to the published version of the manuscript.}}

\funding{{This research was funded by Chinese Scholarship Council.}}
\institutionalreview{{Not applicable}}
\informedconsent{{Not applicable}}
\dataavailability{{Not applicable}}
\acknowledgments{{The authors would like to thank Dr. Tuomas Aittomaki for his valuable suggestions for this paper.}}

\conflictsofinterest{{The authors declare no conflict of interest.}}

\newpage
\abbreviations{Abbreviations}{
The following symbols are used in this manuscript:\\

\noindent 
\begin{tabular}{@{}lp{11cm}<{\raggedright}}
		$\alpha\;\;\;$ &scalar, complex path loss \\
		$\mathbf{A}\;\;\;$ &transmitted signal matrix under block fading assumption\\
		$\mathcal{A}\;\;\;$ &communication user set\\
		$\mathcal{A}_r\;\;\;$ &communication user set that is interfered by radar\\
		$\mathcal{A}_c\;\;\;$ &communication user set that is not interfered by radar\\
		$\beta \;\;\;$ & baseband subcarrier\\
		$\beta_R \;\;\;$& radar baseband subcarrier\\
		$\beta_C \;\;\;$& communication baseband subcarrier\\
		$\mathbf{B}\;\;\;$& {frequency modulation matrix}\\
		$\mathbf{\ddot{B}}\;\;\;$& demodulation matrix\\
		$\mathcal{B}\;\;\;$& radar user set\\
		$\mathbf{C}\;\;\;$& channel coding matrix\\
		$d \;\;\;$ &scalar, number of degrees of freedom\\
		$\mathbf{D}\;\;\;$& interference plus noise covariance matrix\\
		$h[f_1] \;\;\;$& scalar, channel impulse response received for first subcarrier\\ 
		$\mathbf{H}\;\;\;$ &channel matrix\\
		$K \;\;\;$ & total number of communication users\\
		$L \;\;\;$&  total number of blocks/pulses\\
		$M \;\;\;$ & total number of subpulses\\
		$N \;\;\;$ & column dimension of precoder matrix, which  is user-desired frequency DoFs.\\
		$N_{sc}\;\;\;$ & number of transmitted subcarriers\\
		$N_{p}\;\;\;$ & column dimension of data matrix $\mathbf{S}$  \\
		$\mathbf{P}\;\;\;$& $N_{sc} \times N$-dimensional precoding matrix\\
		$\mathbf{Q}\;\;\;$& $N_{sc} \times N$-dimensional decoding matrix\\
		$\mathbf{s}(N)\;\;\;$ & data  vector for $N$th time slot \\
		$\mathbf{S}\;\;\;$& $N_{p} \times M$-dimensional data matrix\\
		$T_p \;\;\;$&  entire pulse duration\\
		$\mathbf{W}\;\;\;$& $N_{sc} \times M$-dimensional {Gaussian noise matrix}\\
		$\mathbf{Y_T}\;\;\;$& $N_{sc} \times M$-dimensional transmitted signal matrix \\
		$\mathbf{Y_R}\;\;\;$& $N_{sc} \times M$-dimensional received signal matrix \\
		$\tau_{T,i} \;\;\;$& scalar, delay between transmitted antenna and target\\
		$\tau_{R,i} \;\;\;$& scalar, delay between received antenna and target\\
		$\theta\;\;\;$&  target's direction of arrival\\
		$\bm \Omega \;\;\;$& selection matrix\\
		$\mathbf{1} \;\;\;$&  {all-ones matrix}\\
\end{tabular}}

\appendixtitles{no} 
\appendixstart
\appendix
\section[\appendixname~\thesection]{}\label{app1}

A multicarrier radar and communication signal model with  $N_{sc}$ available subcarriers may be described as in  \eqref{transmittersignal}. The coded data/waveforms are mapped to subcarriers after precoding and channel coding. Pulse multicarrier radar signals and {continuous multicarrier communication signals are employed in this model.}

\vspace{-6pt}
\begin{figure}[H]
	\normalsize
   	\footnotesize
  \begin{myequation}\label{transmittersignal}
	{Y_{T}}(k) =  \begin{cases}
	2Re\{ \sum_{x=0}^{N_{sc}-1}\sum_{y=0}^{N_p-1}\sum_{m=0}^{M-1} \sum_{n=0}^{N-1} p_{k,n} c_{n,y}s_{y,m}\,
	rect[\frac{kT_c}{N_{sc}}-(x+ a  m)T_c]\,\times e^{j2 \pi (f_c +x\Delta f) \frac{kT_c}{N_{sc}}}
	\,\} &\text{for radar } \\
	2Re\{ \sum_{x=0}^{N_{sc}-1}\sum_{y=0}^{N_p-1}\sum_{m=0}^{M-1} \sum_{n=0}^{N-1} p_{k,n} c_{n,y} s_{y,m}\,
	rect[\frac{kT_c}{N_{sc}}-xT_c]\,e^{j2 \pi (f_c +x\Delta f) \frac{kT_c}{N_{sc}}}
	\}, &\text{for communication }
	\end{cases}
	\end{myequation}
\end{figure}\vspace{-12pt}

A rectangular pulse shape function is denoted by $rect(t)$ in \eqref{transmittersignal}. $Y_T(k)$ is an element of matrix $ \mathbf{Y_T}$. The remaining notations are defined as follows:
\begin{itemize}
	\item  $m=\{0,1,2,\dots,M-1\}$ is the index of subpulses, where $M$ is the number of subpulses.
	\item  $n=\{0,1,2,\dots,N-1\}$ is the index of channel coded data sequence, where $N$ is the number of channel coded data symbols; it essentially defines user-desired DoFs.
	\item  $y=\{0,1,2,\dots,N_p-1\}$ is the index of data sequence before channel coding, where $N_p$ is the number of original, uncoded data symbols.
	\item  $x=\{0,1,2,\dots,N_{sc}-1\}$ denotes the index of subcarriers, where $N_{sc}$ is the total number of transmitted subcarriers.
	\item  $k=\{0,1,2,\dots,MN_{sc}-1\}$ denotes the sample index.
        \item  $a$ is a multiplicity factor between communication symbol duration $T_c$ and
          radar subpulse duration $T_p$, where $T_p$ is larger than $T_c$ {to avoid range aliasing}.
          Then, the radar subpulse duration is given by $T_p=aT_c$.
        \item  $c_{n,y}$ is a scalar with respect to the $n$th subcarrier and $y$th data symbol, $c_{n,y}$ is {an element of coding vector for radar signal} and the channel coding operator in the communication signal; e.g., it will be $\exp(j\phi)$ for P3 or P4 radar code and an element of the Turbo generator matrix for Turbo code.
        \item $s_{y,m}$ is scalar data corresponding to the $y$th data symbol and $m$th subpulse, which is {considered random} for the payload communication signal and known in radar applications because the transmitted waveform information is known by its receiver.
        \item $p_{k,n}$ is a scalar that denotes the precoding coefficient with respect to the $k$th subcarrier and $n$th subpulse before the signal emitted.
\end{itemize}

A detailed derivation of \eqref{transmittersignal} is given in the following. Considering an OFDM signal with $N_{sc}$ subcarriers and $\Delta f$ Hz {subcarrier spacing}, the continuous-time complex envelope of the baseband OFDM signal can be written for any time $t$ as
\begin{equation}
 s_m(t) =  \sum_{x=0}^{N_{sc}-1} w_{m,x}\,rect[t-xT_c]\,e^{j2 \pi x \Delta ft}
\end{equation}
where $rect[\cdot]$ is a discrete rectangular window, $T_c$ is the symbol duration, and the constraint $T_c = \frac{1} {\Delta f}$
satisfies the spectrum selection channel condition. $w_x$ indicates the amplitude weighting
of the $x$th subcarrier. Additionally,
\begin{equation}
rect(t)=
\begin{cases}
1&\text{if \,$0 \leq t \leq T_c$},\\
0&\text{otherwise}.
\end{cases}
\end{equation}

The {transmitted multicarrier} communication signal is considered to be continuous:
\begin{equation}
 y_{T,c}(t) =\sum_{m=-\infty}^{\infty} s_m(t-mT_c).
\end{equation}

In general, a multicarrier radar signal can be written as follows:
\begin{equation}\label{OFDMmultitrans}
 y_{T,r}(t) =\sum_{m=0}^{M-1} s_m(t-mT_p) ,
\end{equation}
where $M$ is the number of OFDM subpulses, $T_p = a T_c$, $a \in \mathbb{R}^+$ is the radar pulse duration.
As both radar and communication signals coexist,  are designed jointly and are postprocessed simultaneously in this paper, {a common clock is shared between two subsystems}. The transmitted signal is a real part of the baseband complex envelope;
then, \eqref{OFDMmultitrans} can be rewritten as~follows:
\begin{equation}
  \begin{split}
 \widehat{y}_{T,r}(t) &=2Re\{ \sum_{m=0}^{M-1}s_m(t-mT_p) \, \}\\
 & =2Re\{ \sum_{m=0}^{M-1}\sum_{x=0}^{N_{sc}-1} w_{m,x}\,rect[t-xT_c-mT_p]\, \}  \\
 &e^{j2 \pi x \Delta ft}\,\} \\
 &=2Re\{ \sum_{m=0}^{M-1} \sum_{x=0}^{N_{sc}-1} w_{m,x}\,rect[t-(x+ a  m)T_c]\, \\
 &e^{j2 \pi x \Delta ft}
 \,\} 
 \end{split}
 \end{equation}
 
According to the Nyquist--Shannon sampling theorem, the sampling rate
is $N_{sc}\Delta f$ samples/second.
Additionally, replacing $  w_{m,x}$ with $p_{k,n}$, $c_{n,y}$ and $s_{y,m}$, the radar transmitted signal can be rewritten as follows:
\begin{equation}
 \begin{split}
\widehat{y}_{T,r}(k) = &2Re\{ \sum_{x=0}^{N_{sc}-1}\sum_{y=0}^{N_p-1}\sum_{m=0}^{M-1} \sum_{n=0}^{N-1} p_{k,n} c_{n,y}s_{y,m}\,\\
&rect[\frac{kT_c}{N_{sc}}-(x+ a  m)T_c]\,\times e^{j2 \pi (f_c +x\Delta f) \frac{kT_c}{N_{sc}}}\}
\end{split}
\end{equation}

The transmitted communication signal can also easily be derived. 

\section{}\label{app2}
From \eqref{eq3}, we define $  \mathbf{Y}_{\mathbf{T}}=(\bm \Omega\circ \mathbf{B})\mathbf{P}
\mathbf{C}\mathbf{S} $ as a compact matrix form of transmitted signal.
The received response in the time domain {is} \cite{bica2016generalized}
\begin{equation}
 \begin{split}
\mathbf{Y_{R}} = \mathbf{\ddot{B}}\mathbf{G}( \bm \theta )
\mathbf{Y}_{\mathbf{T}} 
\end{split}
\end{equation}
where $\mathbf{\ddot{B}}$ is the demodulation matrix, $\mathbf{\ddot{B}}\mathbf{B}= \mathbf{I}_{N_{sc}\times N_{sc}}$.
Considering that the channel coefficients are  {$h_i(\theta), i = 1, \cdots, N_{sc}$, }a guard interval (or cyclic prefix) with length $N_{sc}$ is employed. Note that the signal is transformed into the frequency domain after modulation. To complete the convolution through matrix multiplication,  $\mathbf{G}( \bm \theta)$ is defined as a circulant matrix that is given by the following $N_{sc}$ by $N_{sc}$ matrix~\cite{wang2016overview},
\begin{equation}
 \mathbf{G}( \bm \theta) = \begin{pmatrix}
h_1(\theta)& h_2(\theta)& \cdots & h_{N_{sc}}(\theta)\\
h_{N_{sc}}(\theta)& h_1(\theta)& \cdots & h_{N_{sc}-1}(\theta)\\
\cdots & \cdots & \cdots& \cdots \\
h_2(\theta)& h_3(\theta)& \cdots & h_1(\theta)\\
\end{pmatrix}
\end{equation}

The circulant matrix $\mathbf{G}( \bm \theta)$ can be diagonalized by eigenvalue decomposition. In the OFDM case, $\mathbf{G}( \bm \theta)$ can be written as follows:
\begin{equation}
\mathbf{G}( \bm \theta) = \mathbf{B}^{H} \mathbf{H}( \bm \theta) \mathbf{B}.
\end{equation} 
where  {${\textbf{H}(\bm{\theta})}= \text{diag} \{ h_1(\theta), h_2(\theta) \cdots ,h_{N_{sc}}(\theta)\}$} is a diagonal eigenvalue matrix. Matrix $\mathbf{B}$ is symmetric and unitary for the OFDM signal,
$\mathbf{\ddot{B}}= \mathbf{B}^{H} = \mathbf{B}$. Then, the received signal is given by the following:
\begin{equation}
 \begin{split}
\mathbf{Y_{R}}&= \mathbf{\ddot{B}}\mathbf{G}( \bm \theta ) 
\mathbf{Y}_{\mathbf{T}}  \\
&= \mathbf{B}^{H} \mathbf{B}^{H} \mathbf{H}( \bm \theta) \mathbf{B} \mathbf{B}\circ \bm \Omega\mathbf{P}
\mathbf{C}\mathbf{S}\\
&= \mathbf{H}( \bm \theta ) (\bm \Omega\circ \mathbf{I}) \mathbf{P}
\mathbf{C}\mathbf{S}
\end{split}
\end{equation}

\reftitle{References}

\begin{adjustwidth}{0cm}{0cm}

%


\end{adjustwidth}
\end{document}